# Improving Network Monitoring and Security via Visualization

Keith B. Miller and Eric R. Brandon
Computer Science Department
Utah Valley University, Orem, UT
Email: {keith.miller, eric.brandon}@uvu.edu

*Abstract*—Internet networks are handling increasing volume of traffic than ever before. This data is mainly associated to sensitive, distributed, and multimedia applications. In the past years, much attention has been paid to the way network infrastructure must be designed and developed in order to handle the challenges of delivering high quality services for applications such as VoIP and streaming video.
But the pressures of these sensitive new applications underline a major challenge in network management—how to meet the need for more predictable application delivery despite the fact that IP is simply not predictable. This challenge can't be met by having incrementally more sophisticated ways to understand the flood of existing SNMP management data collected from your network devices. Rather, it requires an expanded network management paradigm that moves from relying only on "point-based" SNMP data to including path and flow awareness through routing and traffic-flow analysis. IT organizations that grasp this shift will have the visibility they need to ensure the success of their inexorable march to a converged network reality.
This paper explores methods of improving network management, monitoring and security through visualization.

## I. Introduction

IP networks are critical infrastructure, transporting application and service traffic that powers productivity and customer revenue. Yet most network operations departments have no way to monitor the IP-layer operation to ensure that the network is able to deliver traffic stably and predictably. The reason is that network monitoring has historically been based on SNMP polling of device and interface status and statistics. While certainly useful, information from individual devices and interfaces can't convey the complex inter-working status of the devices as a whole network. As a result, operations managers are lacking critical monitoring data, particularly in complex IP network topologies that possess high levels of redundancy, and when MPLS VPNs are a major component of the WAN.

Enter route analytics, the network management technology that monitors the network's live routing protocol control plane and uses network-wide routing intelligence to turn sparse amounts of Netflow into network-wide traffic flow visibility. By implementing route analytics, network operations managers responsible for large, complex IP networks can increase the speed and efficiency of network monitoring and reduce operations costs while increasing customer satisfaction.

FIGURE 1. THE DESKTOP VERSION OF VISITREND'S VIC3, DISPLAYS INTERCONNECTED AND CONFIGURABLE VIEWS OF NETWORK SECURITY CONDITIONS. READ MORE ABOUT WHAT THIS SCREENSHOT SHOWS BELOW.

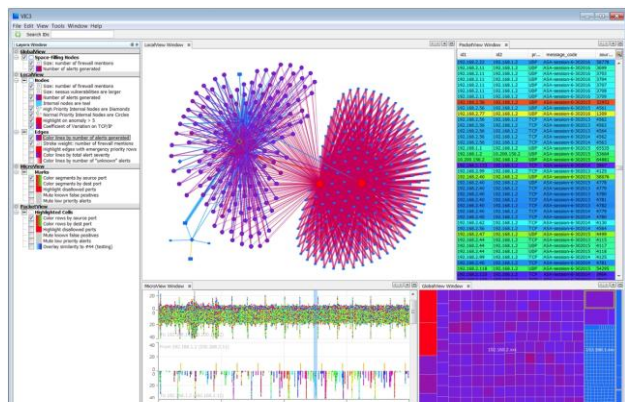

This paper reviews the causes and costs of insufficient network monitoring and related security aspects, explains how route analytics works and how it can be used to enhance network monitoring and troubleshooting by adding real-time visibility into routing operations as well as network-wide traffic flows, leading to operations, engineering and business costs savings.

## II. Related work

There is a significant body of literature in the area of deploying measurement points and studying their characteristics and their impact on improving infrastructure security. Visualization techniques have been also applied to view static data, such as distributed firewall rules to detect potential conflicts or anomalies.
IDMaps [11] studies distance monitoring and estimation by finding distance between Tracers, which are monitoring boxes, placed at various network nodes. The distance maps form the virtual topology of the Internet. However, IDMaps does not assume that each tracer monitors a shortest path tree as it is assumed here.

[8] and [17] study link monitoring and delays in IP networks based on a single point-of-control. PingTV [16] and Atlas [18] uses ICMP to generate a logical map of the network from a single probing host. PingTV monitors the traffic condition and network outages of networks with hierarchical structure by pinging hosts in hierarchical order. Atlas captures the topology of IPv6 networks by probing from an initial set of seeds that grows whenever new routers discovered. Either of these methods is proactive. That is, it sends network probes whereas in our methods we do not require any network probes. In [12] and [13], the authors develop techniques to infer the performance of all of the links (or specified subset of links) that are contained by the trees.

Researchers in [2] and [3] have explored methods to place minimal but comprehensive network defenses automatically by solving graph problems, such as vertex cover and multicuts. While these problems are computationally complex in general, efficient greedy algorithms exist to produce effective solutions.

[15] was the first study to show that the concept of "more measurement points is better" is not accurate by showing that the topology can obtained using few measurement points. [14] studies deploying minimum number of beacons on a network of known topology and BGP-like routing policy so that every link is monitored by messages originating from at least one beacon.

In [1], the authors propose a model to minimize the overhead of monitoring all links of a given network. However, this approach is different from ours as it considers weighted networks and shortest path trees rooted at these nodes are fixed.

PolicyVis [4] is a visualization tool for plotting IP addresses and port numbers specified by the firewall rules. Instead of visualizing policy rules, we visualize the dynamic data, which is the actual network activities made by users' applications. The visual analysis done on the empirical data is a substantial and necessary supplement to the static rules inspection as a proof of correctness to the policy rules. Graph-based network traffic visualization [6] can be used to monitor host behavior, suspicious behavior and network anomalies can be detected through graph drawing [5], graph clustering [7], tree-views on hierarchical clustering (NetADHICT [17]), animated glyphs [18], pixel luminance based histograms (IDGraphs [19]), and scatter plot of IPs/ports in Service Usage Plane in 2D/3D Cartesian coordinates for the flow data [20]. However, lack of the ability for either interactive exploration of context data or intelligent misuse detection modules in these tools make less efficient for insight/knowledge acquisition and root cause identification of various management and security problems. To overcome these constraints, WiNV [21] was developed as dynamic and intelligent visualization tool as opposed to static visualization in that the users can explore the data in a highly interactive manner. Users can click on nodes, perform queries to database on demand, automatically compute and generate statistical charts (e.g., distances), selectively run from a rich pool of algorithms (e.g. clusters evolution), and intelligently guide the administrator to pin down the problem source nodes which require further investigation. Finally, in [22,23] the authors propose capturing the inter-dependencies among network components in 'Leslie graphs,' based on the original dependency work of Lamport. The ''black-box'' approach relies on the correlation of observed network traffic to infer system dependencies. The agents in their system (called AND) perform temporal correlation of the packets sent and received by the hosts; where the central server engine performs Bayesian inference from the reports generated by the agents. While these works mainly focus on computing the dependency graphs for fault localization (i.e. debugging the location of network failure or sluggish performance), our system focuses on the light-weight aspects of information gathering and how to visualize not only connectivity but, the context of the connectivity itself. In short, while these tools help to locate dependency related performance problems at the host-level in a theoretical sense, ENAVis provides a robust platform for exploring and visualizing the connectivity data for a much wider assortment of security and performance-related issues.

## III. THE CAUSE AND COST OF INSUFFICIENT NETWORK MONITORING

IP's distributed routing intelligence makes it efficient and at the same time unpredictable. IP routing protocols automatically calculate traffic routes or paths from any point to any other point in the network based on the latest known state of network elements and network routing configuration. Any changes to those elements cause the routing topology to be recalculated dynamically. While this means highly reliable traffic delivery with low administrative overhead, it also creates endless variability in the active routing topology. Not only can a large network be in any one of millions of possible active routing topology states, but application traffic patterns are by nature unpredictable. Network problems – router software bugs, misconfigurations, hardware that fails (often after exhibiting intermittent instability) – can add to that unpredictability. Unfortunately, traditional SNMP network monitoring tools that operations departments rely on to detect and troubleshoot network problems simply can't perceive the dynamic changes in routing and traffic because they monitor the network on the basis of device and interface status. In a simple network where there is no redundant WAN links, MPLS VPNs or other complex topology, device status does effectively correlate to network status because there is no variation in the way that traffic can possibly transit the network. However, in a redundant or complex network topology, traffic can transit different routers based on the state of Layer 3 routing, which chooses which paths and links traffic uses to get from any point A to any point B. The result from a SNMP monitoring and operations point of view is that when a problem is occurring, it can be very difficult to figure out where exactly in the network to look—which links should be examined to see if they have problems? Furthermore, what if the problem has nothing to do with an interface or device having a hardware problem—what if the routing control plane itself is having a problem such as unstable route advertisement that flaps up and down, causing intermittent reachability issues over time? SNMP solutions simply have no visibility into those "software" problems in the network.

The visibility problem gets even worse when the problem is no longer currently occurring, and is handed off to network engineering. Highly trained, expensive network engineers are reduced to playing a glorified guessing game since they have no forensic audit trail of network routing and traffic conditions at the time the problem occurred. One network engineer at a large regional North America bank called this phenomenon "footprints in the sand"--by the time a problem is being looked at by knowledgeable engineers, all evidence has been washed away by the figurative waves of changing network conditions.

All of this fumbling around in the dark and guessing is expensive to network operations and engineering. The time drain itself costs a lot of productivity within the IT department—this isn't trivial since personnel costs are typically one of, if not the biggest piece of the network operations and engineering budget. More importantly, unsolved problems or delayed resolutions cost endusers the productivity that applications and services are supposed to be delivering. The costs of application downtime are well understood for large organizations—ranging from tens of thousands to millions of dollars per hour, depending on the industry.

## IV. ROUTE ANALYTICS—COST SAVINGS THROUGH GREATER VISIBILITY

Network management's purpose is to overcome the complexity inherent in a large network and automate the work of network operations and engineering personnel so that applications can be delivered with high reliability. While traditional network monitoring approaches aren't sufficient, organizations grappling with the challenges of managing complex network topologies have an answer in route analytics technology.

Route analytics technology works by utilizing the network's live routing protocols as a new source of network management information and intelligence, complementing traditional SNMP data. A route analytics device –a network appliance running specialized software – acts like a router, listening to routing protocol updates sent by all routers in the network, and computing the network-wide routing state in real-time, just as all the "real" routers do. While the route analytics device itself is passive, never advertising itself as a place to send traffic, it provides real-time visibility, always up-to-date routing-state knowledge, and a completely accurate historical record of all past routing changes. It knows every route or "path" that any traffic takes at any point in time – hence the name "path-based" network management. The network-wide routing topology understanding and the full detail of routing changes provides the basis for many useful analyses of the routing control plane. When combined with Netflow traffic-flow data, the full power of route analytics information emerges. By collecting traffic flows from the ingress points of traffic at the network edge (data centers, Internet peerings, and major WAN links), then mapping them to the precise routes they traverse through the network produces an integrated, always accurate map of all routing and traffic for the entire network core.

Integrated routing and Netflow monitoring enables far more efficient network operations processes, leading to better application delivery. Route analytics can monitor a variety of important network conditions that aren't visible to any other network management technology, such as:

- Internal IP subnets, for which network reachability is managed by routing protocols such as OSPF, EIGRP, IS-IS: Route analytics can monitor the availability and stability of important individual subnets, such as those hosting server farms in data centers.
- External Internet networks, managed by the BGP protocol: Route analytics can monitor the availability of subnets on the Internet. For those organizations that maintain multiple Internet peerings to ensure availability of Internet traffic, route analytics can also monitor the level of redundancy of paths to critical Internet networks and servers.
- MPLS VPN-reachable subnets, managed by the BGP protocol: MPLS VPNs typically rob network managers and operators of important visibility, since the routing across the WAN backbone is outsourced to the VPN service provider. Route analytics restores routing and traffic visibility when MPLS VPNs are implemented, monitoring whether all sites can reach each other's networks, traffic levels between sites, and anomalous changes in routing reachability that can occur due to internal or SP errors.
- Layer 3 paths—the specific set of links that traffic must traverse from any point A to any point B in a network. Route analytics can monitor network paths for applications that are sensitive to changes in network latency, or for cases where network paths must stay stable to ensure proper security or other network policies
- "Software"-related link state changes, such as link flapping caused by misconfigurations of routing protocols, network design errors, or router bugs, which can lead to a loss of network or application availability
- Traffic utilization on all links in the network, broken out by CoS or application, without needing to turn on overhead-inducing Netflow export on all the interfaces in the network

## V. CONCLUSION

Network security managers are routinely being called upon to deliver more with less. Route analytics' increased monitoring visibility, forensic history and forward-looking modeling capabilities make it a force-multiplier for network operations and engineering teams, helping network managers ensure the reliable delivery of critical applications and services that drive top and bottom line business results, while containing operational costs.